\documentclass[12pt]{article}

\usepackage{graphicx}
\usepackage{amssymb}
\usepackage{epstopdf}
\usepackage{authblk}
\usepackage{dsfont}

\usepackage{graphicx,epsfig}
\usepackage{graphics,dcolumn,bm,epic, eepic,fleqn,float}
\usepackage{amssymb,amsmath,amsfonts,multirow,rotate,color}
\usepackage{soul}
\usepackage[hyphens]{url}

\definecolor{Myorange}{cmyk}{0,0.42,1,0}

\title{Food-bridging: a new network construction to unveil the principles of cooking}
\author[1,2]{Tiago Simas\thanks{Email: tiago.simas@telefonica.com}}
\author[1]{Michal Ficek}
\author[3,4]{Albert Diaz-Guilera}
\author[1]{\\Pere Obrador}
\author[1]{Pablo R. Rodriguez}
\affil[1]{Telefonica Research, Edificio Telefonica, Barcelona, Spain}
\affil[2]{Department of Psychiatry, University of Cambridge, Cambridge, UK}
\affil[3]{Departament de Fisica de la Materia Condensada, Marti i Franques 1, Universitat de Barcelona, 08028 Barcelona, Spain}
\affil[4]{Universitat de Barcelona Institute of Complex Systems (UBICS), Universitat de Barcelona, Barcelona, Spain}

\begin{document}
\maketitle
\begin{abstract}
In this manuscript we propose, analyse, and discuss a possible new principle behind traditional cuisine: the Food-bridging hypothesis and its comparison with the food-pairing hypothesis using the same dataset and graphical models employed in the food-pairing study by Ahn et al. \cite{YY2011}. 

The Food-bridging hypothesis assumes that if two ingredients do not share a strong molecular or empirical affinity, they may become affine through a chain of pairwise affinities. That is, in a graphical model as employed by Ahn et al., a chain represents a path that joints the two ingredients, the shortest path represents the strongest pairwise chain of affinities between the two ingredients. 

Food-pairing and Food-bridging are different hypotheses that may describe possible mechanisms behind the recipes of traditional cuisines. Food-pairing intensifies flavour by mixing ingredients in a recipe with similar chemical compounds, and food-bridging smoothes contrast between ingredients. Both food-pairing and food-bridging are observed in traditional cuisines, as shown in this work.

We observed four classes of cuisines according to food-pairing and food-bridging: (1) East Asian cuisines, at one extreme, tend to avoid food-pairing as well as food-bridging; and (4) Latin American cuisines, at the other extreme, follow both principles. For the two middle classes: (2) Southeastern Asian cuisines, avoid food-pairing and follow food-bridging; and (3) Western cuisines, follow food-pairing and avoid food-bridging.
\end{abstract}

\section{Introduction}
Since the introduction by Francois Benzi and Heston Blumenthal of the Food-pairing hypothesis, a debate on this hypothesis has been risen in gastronomy science and cuisine. Originally it states that, if two ingredients share important flavour compounds\footnote{By flavour compounds we mean molecular chemical compounds and from now on will describe these as flavours compounds or, in short, flavours.}, there is a good chance that they will result in a tasty combination, \cite{YY2011, YY2013}. In the last few years, this hypothesis attracted foodies, many chefs, and scientists. If food-pairing is one of the main principles behind our taste preferences, scientifically this would allow us to predict and build many successful new ingredient affinities based on which flavours they are composed.

Ahn et al. \cite{YY2011}, present a study of the food-pairing hypothesis across several regional cuisines, employing a set of tools which derive from a new scientific field: complex networks, a sub-field of complex systems \cite{BLMCH2006}. These mathematical techniques were applied to several regional cuisines, to encode a relation between ingredients and flavours as a network, where nodes and edges represent ingredients and flavours sharing respectively. This network is called flavour network from now on. The authors have observed that Western cuisines show a tendency towards the food-pairing hypothesis, i.e. their flavour network contains many pairs of ingredients that share many flavour compounds, and that Eastern Asian cuisines tend to avoid compound sharing between ingredients. The study of food-pairing has lately been applied to other specific regional cuisines \cite{ARG2015,SJMYP2015,VVWM2013}.

The hypothesis of food-bridging stems from the combination of the theory of complex networks and gastronomy \cite{YY2011,YY2013,Erik,ARG2015,fenaroli,SR,SimasTh,SR2012}. It assumes that if two ingredients do not share a strong molecular or empirical affinity, they may become affine through a chain of pairwise affinities. That is, apricot and whiskey gum may not be affine, but if we join (or bridge) them with tomato they may become affine -- assuming that tomato is affine with apricot and whiskey gum, thus creating a chain of affinities. In a graphical model of a flavour network this corresponds to a path that joints the two ingredients, but not necessarily the shortest path. However, the shortest path represents the strongest pairwise chain of affinities between the two ingredients.

In this manuscript we analyse and discuss the food-bridging hypothesis with a restriction to the optimal case, which corresponds to the shortest path in the graphical model. We use the same dataset and graphical models employed in the study of Ahn et al. \cite{YY2011}. This allows us to perform a direct comparative study between food-pairing and food-bridging.

\section{Methods and Materials}

\subsection{Data and Ingredient Networks}
\label{data}
The data used in this work as well as the methods employed to build ingredient networks, are based on the work of Ahn et. al. \cite{YY2011}. In short, the flavour network is a weighted graph obtained from a bi-partite graph that relates $1530$ ingredients with $1106$ flavour compounds \cite{YY2011}. Nodes in the flavour network represent the ingredients, edge weights are the number of flavours compounds shared between pairs of ingredients \cite{YY2011}.

We removed some regional cuisines from the original work \cite{YY2011}. The reason behind is that we employed the null-model Frequency-conserving described in the supplementary materials \cite{YY2011}, and after a permutation test and multi-comparison correction (False Discovery Rate), they show evidence of no statistical difference when compared with the null-model on the variables: food-pairing and food-bridging.

In general, the weights of a weighted network lie in a non-normalised interval $Z_{ij}\in[a,b]\subset\mathds{R}$. Normalising the network weights to the unit interval $I=[0,1]$ does not affect network properties, if the normalisation is performed by a linear function. As shown in \cite{SR} there is only one unique linear function that performs such normalisation.

\begin{equation}
\label{linearnorm}
w_{ij}=\frac{(1-2\epsilon) Z_{ij}+(2\epsilon-1)\cdot MIN(Z_{ij})}{MAX(Z_{ij})-MIN(Z_{ij})}+\epsilon
\end{equation}

\noindent We have parameterise this function with $\epsilon$ in order to avoid merging and/or isolating vertices with weights at the boundaries of $Z_{ij}\in[a,b]$. In general $\epsilon$ is set to $0.01$. 

This normalisation allows us to apply the framework described in \cite{SR}, i.e. allows us to treat weighted graphs as mathematical objects defined in a specific algebra, similar to the way in which we use algebras to deal with numbers.

\subsection{Food-pairing, Food-bridging and Flavour Network Semi-metricity}

\begin{figure}[h!]
\centering
\includegraphics[scale=0.3]{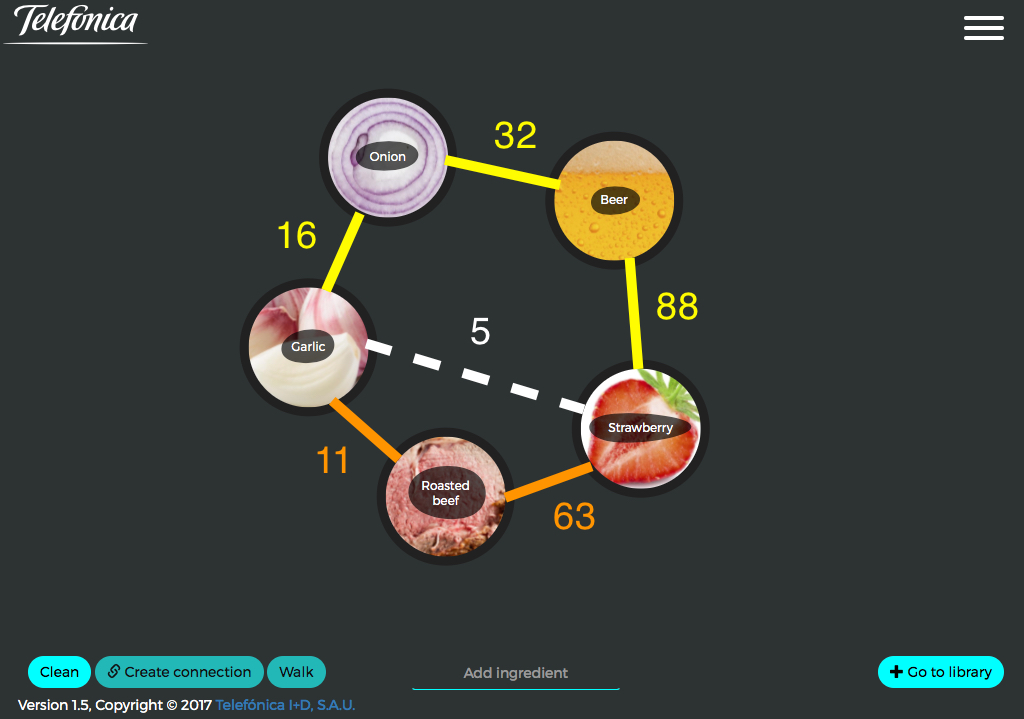}
\caption{Semi-metric edge and paths example from the flavour network. The white edge that links "garlic + strawberry" shares 5 flavours. Therefore it is semi-metric, because there are two alternative paths the yellow and orange paths that better overlap flavours or food-pairs based in a chain of other ingredients, smoothing the initial flavour contrast between these ingredients. The two semi-metric paths that connect "garlic-strawberry" are:  (1) Yellow path, with "garlic + roasted onion + bantu beer + strawberry", which shares 16+32+88 = 136 flavours; and (2) orange path, with "garlic + roasted beef + strawberry", which shares 11+63 = 74 flavours. The first semi-metric path may inspire a "garlic-strawberry" sauce, based on "garlic + roasted onion + bantu beer + strawberry", and with the second semi-metric path may inspire the dish composed of  "roasted beef" with garlic-strawberry sauce. This figure is an adapted print screen of the application developed by Telefonica I+D Appetit Team and available online at http://appetit.lab.tid.es - with Telefonica I+D printing permission.  \label{fig_0}}
\end{figure}

\textbf{\emph{Food-pairing}}. As defined in \cite{YY2011} Food-pairing is measured by the number of flavours a pair of ingredients share. The food-pairing value of a recipe is the average number of shared flavours in the recipe, as defined in \cite{YY2011} and is calculated from the following equation:

\begin{equation}
\label{eq_fp}
N_s(R)=\frac{2}{n_R(n_R-1)}\sum_{i,j\in R,i\neq j} |C_i\cap C_j|
\end{equation}

\noindent where $C_k$ corresponds to the edge weight between the pair of ingredients in the flavour network, and $n_R$ is the number of ingredients in the recipe $R$. Each recipe defines a sub-graph in the flavour network and $N_s(R)$ corresponds to the average of all edges in such sub-graph.
\\ \\
\textbf{\noindent\emph{Metric and Semi-metric edges and paths}}. As defined in \cite{SR2012,SimasTh,SR,Simas2015,Simas2016,KSL,SCRG}, an edge in a weighted graph is metric if the shortest path is equal to the edge by itself (direct connection). Otherwise the edge is considered semi-metric, which means that there is at least one alternative path that involves other nodes. See figure \ref{fig_0} for an example.

We may observe in a network of ingredients that two nodes are more strongly connected by other paths (semi-metric paths), whether or not there is a direct edge between them. Figure \ref{fig_0} shows an example of the combination of "garlic" and "strawberry" from the flavour network, which share $5$ flavours when mixed together. In this figure we show how we may increase the poor affinity between these two ingredients by adding additional ingredients that play in the semi-metric paths of the flavour network. From the flavour network, at least two semi-metric paths are stronger than the edge that connects them. In this case, among the possible stronger paths, the optimal semi-metric path is the path that indirectly connects the two ingredients in this network; that is, the path "garlic + roasted onion + bantu beer + strawberry". These intermediate ingredients potentiate the affinity between  "garlic - strawberry". Other semi-metric paths may exist as we show in this example: "garlic + roast beef + strawberry".

Food-pairing is a particular case for which we only consider direct connections, if they exist. In another words $k=0$ $hops$ (zero nodes in between). However, with semi-metric paths we allow two ingredients to be strongly connected with $k>0$ $hops$, whether the edge between the ingredients exists or not.

There are many ways to calculate such alternative paths between any two nodes in a weighted graph. Refer to the Figure 1 in \cite{Simas2016} for a summary of this calculation and, see further details in \cite{SR,Simas2015,Simas2016,KSL,JOHNSON}.
\\ \\
\textbf{\noindent\emph{Food-bridging}}. Food-briging is a hypothesis which assumes that if two ingredients do not share a strong molecular or empirical affinity, they may become affine through a chain of pairwise affinities. That is, apricot and whiskey gum may not be affine, but if we join (or bridge) them with tomato they may become affine -- assuming that tomato is affine with both apricot and whiskey gum, thus creating a chain of affinities.
\\ \\
Food-bridging is the ability to connect a pair of ingredients, that may or may not have a direct connection, through a path of non-repeating ingredients within a network of ingredient affinities; in the specific case of this work the flavour network. Several paths may exist, or none. In the case that no path exists, we say that no bridge exists, otherwise a bridge exists and all possible bridges are ranked by the strength of the path.
\\ \\
\noindent The notion of food-bridging is directly related to semi-metric connections between ingredients in a network of ingredients as briefly mentioned above. Semi-metricity in weighted graphs is a mathematical property of distance that measures all levels of triangle inequality violations. That is, all $k-hop$ inequalities violations between two ingredients, where $k\geq1$ means that we have $k$ intermediate ingredients, see Figures \ref{fig_0} and \ref{fig_2a}(B) for an example.The degree of food-bridging in a recipe is defined as an average of all semi-metric edges in a recipe, or by the average strength of all optimal semi-metric paths, respectively.
\\ \\
More specifically, we define the recipe optimal food-bridging strength $N_{sm}^{*}(R)$ as an average of the strengths of all the optimal paths between any pair of ingredients in the defined recipe sub-graph:

\begin{equation}
\label{eq_fb}
N_{sm}^{*}(R)=\frac{2}{n_R(n_R-1)}\sum_{i,j\in R,i\neq j} \frac{\delta(s_{i,j}>1 \land s_{i,j}<+\infty)}{dc_{i,j}+1}
\end{equation}

\noindent where $$s_{i,j}=\frac{d_{i,j}}{dc_{i,j}}$$ is the semi-metric ratio in the weighted sub-graph defined by the recipe $R$ in the flavour network  \cite{SR,Simas2015,Simas2016,KSL}: $d_{i,j}$ is the direct distance and $dc_{i,j}$ denotes the shortest path between ingredients $i$ and $j$, respectively. $\delta$ is the \emph{discrete-Kronecker} function, i.e. $\delta(condition)=1$, if the logical condition $True$, otherwise $\delta(condition)=0$ -- for logical condition equals False. That is, the equation numerator; $\sum_{i,j} \delta(s_{i,j}>1 \land s_{i,j}<+\infty)$, counts only the semi-metric edges.

Although we define the recipe optimal food-bridging strength $N_{sm}^{*}(R)$, in this work we measure the degree of food-bridging in a recipe as the average of all semi-metric edges in a recipe, which represents a simpler version.
\\ \\
\textbf{\noindent\emph{Network semi-metric percentage}}. As defined in \cite{SR,Simas2015,Simas2016,KSL} the network semi-metric percentage is given by the following equation:

\begin{equation}
\label{eq_smp}
SMP=\frac{\sum_{i,j} \delta(s_{i,j}>1 \land s_{i,j}<+\infty)}{\sum_{i,j} \delta(s_{i,j}\geq1 \land s_{i,j}<+\infty)}
\end{equation}

\noindent where $s_{i,j}$ is the semi-metric ratio between ingredients $i$ and $j$ in the flavour network. The dominator of this equation; $\sum_{i,j} \delta(s_{i,j}\geq1 \land s_{i,j}<+\infty)$ counts all edges in the network and the numerator; $\sum_{i,j} \delta(s_{i,j}>1 \land s_{i,j}<+\infty)$, counts only the semi-metric edges.
\\ \\
\textbf{\noindent\emph{Recipe food-bridging percentage}}. Semi-metric percentage of the sub-graph representing a recipe in the ingredient network is called recipe food-bridging percentage. In other words, Equation \ref{eq_smp} above is applied to the sub-graph defined by the recipe.
\\ \\
\textbf{\noindent\emph{Network metric backbone}}. As defined in \cite{SR,Simas2015,Simas2016,KSL} the metric backbone is the smallest weighted sub-graph of a weighed graph that preserves the shortest paths: sub-graph with all metric edges.
\\ \\
\textbf{\noindent\emph{Network semi-metric backbone}}. The semi-metric backbone is a sub-graph of a weighted graph with only semi-metric edges, i.e. all metric edges removed from the network \cite{Simas2015,Simas2016}.

\section{Results}
\begin{figure}[h!]
\centering
\includegraphics[scale=0.12]{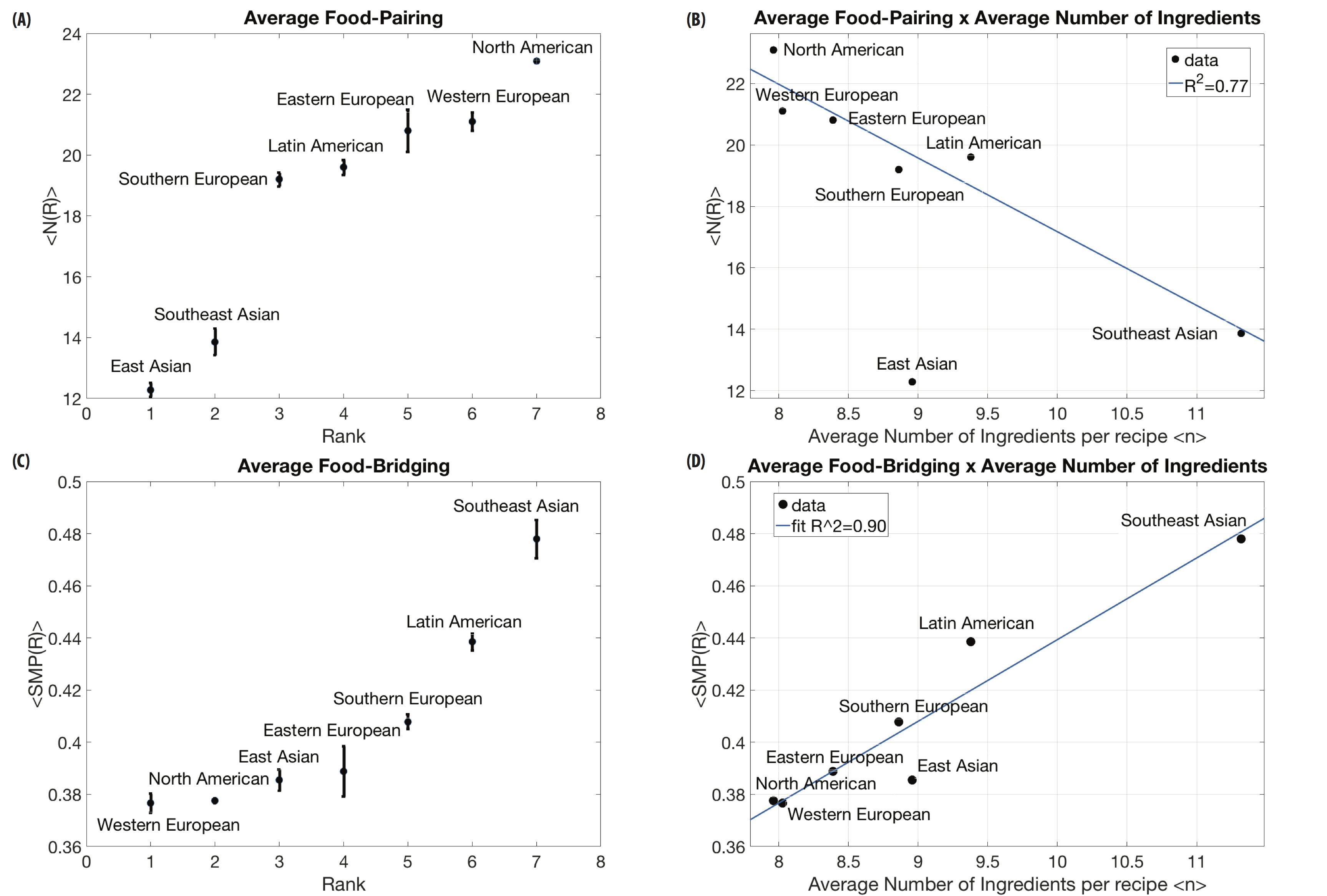}
\caption{$<N(R)>$ is the average over all recipes of the recipe food-pairing, $<SMP(R)>$ is the average over all recipes of the recipe food-bridging, Rank is the sorted cuisine type by $<N(R)>$ or $<SMP(R)>$, and $<n>$ the average number of ingredients per recipe for a given cuisine type. (A) Average recipe Food-pairing ($<N(R)>$)  vs. cuisine type Rank. (B) Average recipe Food-pairing ($<N(R)>$) vs. average number of ingredients per recipe ($<n>$) . (C) Average recipe Food-bridging ($<SMP(R)>$) vs. cuisine type Rank. (D) Average Food-bridging ($<SMP(R)>$) vs. average number of ingredients per recipe ($<n>$).\label{fig_1}}
\end{figure}

In Figure \ref{fig_1} we analysed food-pairing and optimal food-bridging (semi-metric percentage) according to equations \ref{eq_fp} and \ref{eq_smp}, respectively. We plotted the averages of these variables for each of the seven distinct world regions, against how they rank, (Figure \ref{fig_1} (A) and (C)), and against the number of ingredients, (Figure \ref{fig_1} (B) and (D)).
\\ \\
\noindent\textbf{\emph{Food-pairing}:}
In Figures \ref{fig_1} (A) and (B) we observe that there are clearly two distinct groups with respect to food-pairing: Western-based cuisines; and Eastern Asian cuisines. It corroborates with the observations in \cite{YY2011,YY2013,ARG2015} that Eastern Asian cuisines avoid food-pairing more than the Western based cuisines. Moreover, from Figure \ref{fig_1} (B), we observe that there is a negative trend of food-pairing against the average number of ingredients used in a recipe. Note that in this case, East Asian and Southeast Asian cuisines differ mainly in the average number of ingredients used in a recipe. East Asian cuisine is a complete outlier in this trend -- flagging that Southeast Asian cuisine may differ from East Asian cuisine in some other dimension. We also observe that Eastern European as well as Southeast Asian cuisines show higher variability, suggesting a richer cuisine. In fact, the source of the variability may stem from a size effect, since these two cuisines present lower volume of recipes when compared to the others, a collection of 381 and 457 recipes, respectively,  with the others containing over 2000 recipes each.
\\ \\

\noindent\textbf{\emph{Food-bridging}:}
In Figures \ref{fig_1} (C) and (D) we observe that food-briding ranks non-linear and depends positively linearly on the average number of the ingredients used in recipes. From the rank we observe that, in this case, food-bridging: East and Southeast Asian cuisines differ significantly from each other; Western cuisines cluster together at the bottom extreme; and Southeast Asian at the other extreme.

In this case we note that food-bridging (semi-metric percentage) depends on the number of ingredients used in a recipe. This was an expected result since there are more possibilities to bridge ingredients, i.e. more degrees of freedom.
\\ \\

\begin{figure}[h!]
\centering
\includegraphics[scale=0.17]{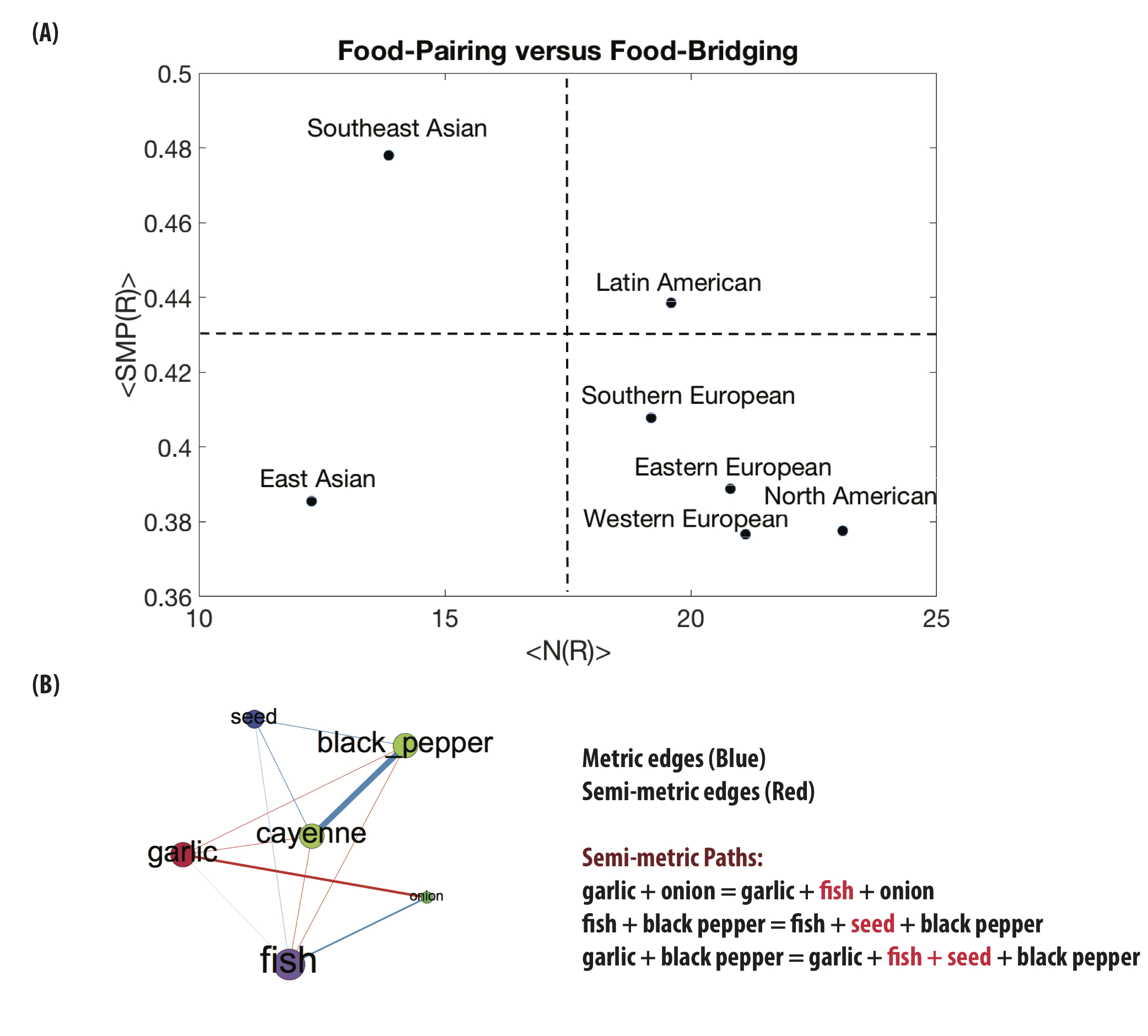}
\caption{$<N(R)>$ is the average over all recipes of the recipe food-pairing, and $<SMP(R)>$ is the average over all recipes of the recipe food-bridging. (A) Classes of cooking based on average recipe Food-pairing ($<N(R)>$) versus average recipe Food-bridging ($<SMP(R)>$) . We divided the space into four distinct regions or classes, as follows: (1) \textbf{Low} food-pairing + \textbf{Low} food-bridging, (2) \textbf{Low} food-pairing + \textbf{High} food-bridging, (3) \textbf{High} food-pairing + \textbf{Low} food-bridging, (4) \textbf{High} food-pairing + \textbf{High} food-bridging. (B) An example of semi-metric analysis of a Southeast Asian cuisine with six ingredients. The semi-metric percentage of this recipe is $SMP=\frac{5}{5+6}\times100\%=45\%$. It shares in average eleven flavour compounds between pairs of ingredients, falling into class (2) according to (A). Moreover, this recipe has nine possible semi-metric paths or bridges, of which some are shown above.\label{fig_2a}}
\end{figure}

\noindent\textbf{\emph{Food-pairing and Food-bridging}:}
In Figure \ref{fig_2a} (A) we observe the relation between food-pairing and food-bridging. We divided the group into four non-overlapping regions, which represent the following four classes.

\begin{enumerate}
\item \textbf{Low} food-pairing + \textbf{Low} food-bridging.
\item \textbf{Low} food-pairing + \textbf{High} food-bridging.
\item \textbf{High} food-pairing + \textbf{Low} food-bridging.
\item \textbf{High} food-pairing + \textbf{High} food-bridging.
\end{enumerate}

We observe that East Asian falls into class (1), Southeast Asian into class (2), Southern, Eastern, Western European and North American into class (3) and Latin American into class (4).

In class (1), with low food-pairing and low food-bridging, the recipe ingredients depend less on the co-occurrence of their flavour compounds, directly or indirectly (chains of pairings). Class (2) pairs flavour compounds mainly indirectly by chains or bridges between ingredients. In class (3), the ingredients mainly pair their flavour compounds without that many indirect chains or bridges. In class (4) the ingredients strongly pair and bridge their flavour compounds.

In Figure \ref{fig_2a} (B) we have an example of a SouthEast Asian recipe with six ingredients\footnote{The meaning of general ingredients products from Fenaroli's book of Flavors \cite{fenaroli} is for example:\\ \\\textbf{fish}: sweet fish, fatty fish, raw fatty fish, ...
\\ \\
\textbf{seed}: lovage seed, toasted sesame seed, angelica seed, ...}. We can observe five semi-metric edges (red) and six metric edges (blue). The semi-metric percentage of this recipe is $SMP=\frac{5}{5+6}\times100\%=45\%$. It shares in average eleven flavour compounds between pairs of ingredients, falling into class (2) according to figure \ref{fig_2a} (A). Moreover, this recipe has nine possible semi-metric paths or bridges, where some of them are shown in the Figure \ref{fig_2a} (B).

\begin{figure}[h!]
\centering
\includegraphics[scale=0.35]{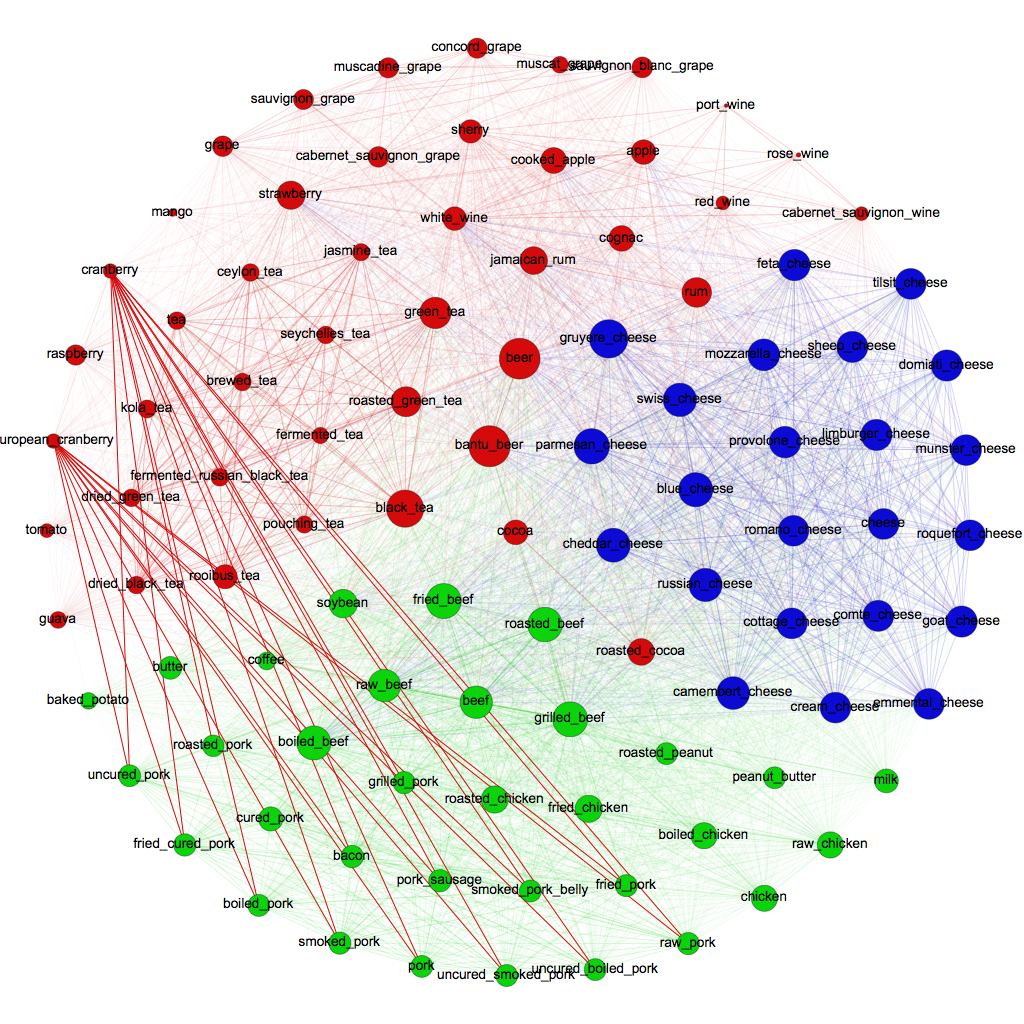}
\caption{\textbf{Metric Backbone}. Top 100 ingredients of Flavour Network with higher node strength.  Edges represent metric connections and edge colour the target community colour (target node). Node colours represent network clusters after applying a community detection algorithm, e.g. Louvain algorithm, and size proportional to the node metric strength. \label{fig_2b1}}
\end{figure}

\begin{figure}[h!]
\centering
\includegraphics[scale=0.35]{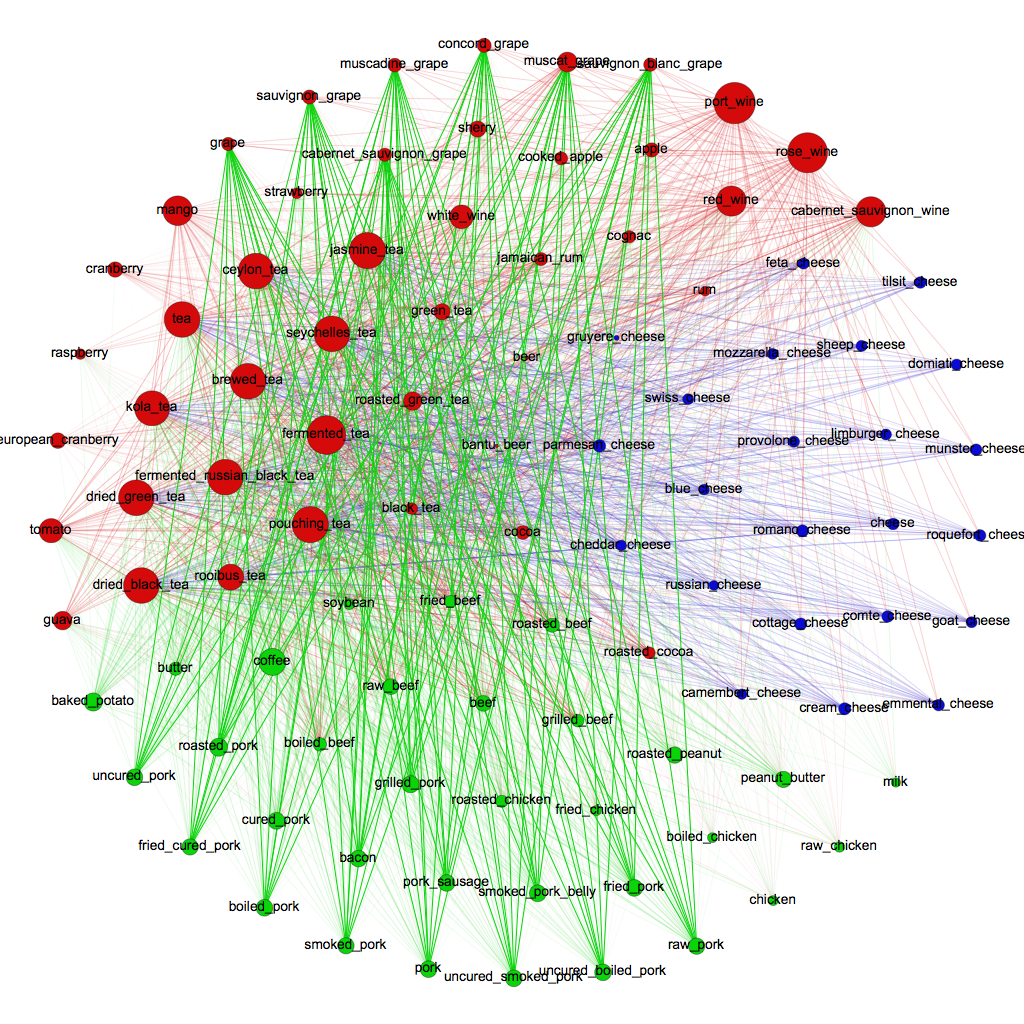}
\caption{\textbf{Semi-metric Backbone}. Top 100 ingredients of Flavour Network with higher node strength. Edges represent semi-metric connections and edge colour the target community colour (target node). Node colours represent network clusters after applying a community detection algorithm, e.g. Louvain algorithm, and size proportional to the node semi-metric strength. \label{fig_2b2}}
\end{figure}

In Figures \ref{fig_2b1}-\ref{fig_2b2}, we show a sub-graph of the flavour network with the top 100 ingredients that have stronger connections or pairings (node strength). Figure \ref{fig_2b1} edges represent only metric connections (metric backbone) and Figure \ref{fig_2b2} edges show only semi-metric connections (semi-metric backbone). Node colours represent network clusters after applying a community detection algorithm, e.g. Louvain algorithm \cite{fortunato2010}, and node size proportional to the node metric or semi-metric strength, respectively. The metric percentage is $27,4\%$ of the edges from the flavour network and the semi-metric percentage is $72,6\%$ from the flavour network, which demonstrates that there are a high number of bridge possibilities between pairs of ingredients.

Highly metric ingredients (node size) tend to food-pair in pairs, and highly semi-metric ingredients (node size) tend to food-pair with the addition of intermediate ingredients. For example, from Figure \ref{fig_2b1} the ingredients "beer", "black tea", "gruyere cheese" etc, are good food-pairing ingredients. Figure \ref{fig_2b2} shows that "port wine", "rose wine", "tea", "tomato" are better mixed with intermediate ingredients, according to food-bridging hypothesis.

In general, we may observe from Figures \ref{fig_2b1}-\ref{fig_2b2} that there is a dichotomy; with ingredients that are less suited to food-pairing tending to use the food-bridging mechanism, and vice-versa.

\section{Discussion}
We have shown that the flavour network is $72,6\%$ semi-metric, which allows food-bridging to work extensively, i.e. the number of possible semi-metric paths between ingredients is large. Food-bridging or semi-metricity, by hypothesis may increase affinities between ingredients with or without a strong direct affinity based on chain of intermediate ingredient pairings, in this case a chain of flavour compounds affines.

Food-pairing and food-bridging are different hypothesis that may represent possible mechanisms behind traditional cuisines. Food-pairing intensifies flavour with similar flavoured ingredients and food-bridging smooths contrasted flavoured ingredients in a recipe, respectively. Both hypothesis food-pairing and food-bridging, are jointly observed in traditional cuisines, as shown in this work.

Regional cuisines cluster in four distinct classes  defined by the possible relationship between these two mechanisms. Where East Asian cuisine is at one extreme class (1), and tends to avoid food-pairing as well as food-bridging; and Latin American cuisine is at the other extreme class (4), following both principles. Southeastern Asian and Western cuisines are in middle classes (2) and (3): class (2) avoids food-pairing and follows food-bridging; class (3) follows food-pairing and avoids food-bridging.

It is worthwhile noting from figure \ref{fig_2a} (A) that the represented cuisine classes follow in some way their geo-political distribution.

East Asia cuisine, represented by Korean, Chinese, and Japanese cuisines, tends to use contrasted ingredients with respect to flavour. This results in a cuisine that contrasts several flavours.

At the other extreme is Latin American cuisine, represented by Caribbean, Central America, South American, and Mexican cuisines. These tend to reinforce the intensity of flavour using both mechanisms, food-pairing and food-bridging. That is, direct and indirect intensification of flavours in a recipe, reinforcing common flavours and smoothing contrasts between flavoured contrasted ingredients.

In class (2) we found Southeast Asian cuisine, represented by Indonesian, Malaysian, Filipino, Thai, and Vietnamese. These cuisines are similar to East Asian cuisines with respect to food-pairing, using contrasted ingredients, but at the same time they smooth these contrasts with other ingredients that bridge the contrast.

The other intermediate class is class (3), where we found Eastern, Southern, Western European, and North American cuisines. Eastern European cuisines are represented by Eastern Europe in general, and Russian cuisines. Southern European cuisine is represented by Greek, Italian, Mediterranean, Spanish, and Portuguese. Western European cuisine is represented by French, Austrian, Belgian, English, Scottish, Dutch, Swiss, German, and Irish. And North American is represented by American in general, Canada, Cajun, Southern soul food, and Southwestern U.S. These cuisines tend to follow the food pairing with the direct intensification of flavours in a recipe, avoiding contrasted ingredients. Therefore, these cuisines are characterised by avoiding flavour contrasted ingredients. Moreover, in this class, at one extreme we have Southern European, and at the other North American. The latter sub-clustering better with Western and Eastern European cuisines. 

We may suggest several explanations for why, in this analysis, traditional cuisines cluster in this way across these two dimensions: food-pairing and food-bridging. The clustering aligns well with a geo-political distribution. These cuisines may be driven by particular geographical weather and resource constrains as well as political trade in goods, which may influence the different styles of cuisine analysed in this work. 

Food-bridging, as shown, opens the possibility of better understanding possible mechanisms behind mixing ingredients in a recipe. This is a new mechanism or hypothesis, different from food-pairing, and both mechanisms are observed in traditional cuisines, in particular in this dataset \cite{YY2011}.

We recognise a number of limitations in this work. We have not included in this analysis important features such as texture, ingredient concentrations, processes used during the recipes, such as cooking method among others \cite{Erik}. We restricted our analysis to the number of shared chemical flavour compounds between ingredients as in the works of Ahn et al. \cite{YY2011,YY2013}. However, for food-bridging, a contra-part of its mathematical representation -- semi-metricity -- is not restricted to the flavour space or more specifically to the flavour network. In general, it may be employed to other modalities: texture, colour, among other empirical or scientifically affinities.

Besides this work, semi-metricity as a topological property of weighted graphs has been shown to be a topological analysis, sensitive and specific in identifying how the flow of information propagates in the human brain \cite{Simas2015,Simas2016}, provide better recommendations in social networks \cite{SR,SR2012}, and a better optimisation of large scale graphical algorithms \cite{KSL}.

This work brings a new perspective on food-pairing, and introduces food-bridging as a new principle or vector behind cooking.

\section*{Author Contributions}
All authors made significant contributions to the drafting of the article.

\section*{Acknowledgements}
The authors would like to acknowledge Telefonica I+D for all support done in the Appetit project that have turn possible the present work. We also thank all of our colleagues involved in the Appetit project. We would like to acknowledge Oliver Smith and Emily Stott for their persistence on editing this manuscript.

\end{document}